\documentclass[conference,a4paper]{IEEEtran}
\usepackage[T1]{fontenc}
\usepackage[utf8]{inputenc}
\usepackage{cite,amsmath,amssymb,graphicx,array,tabularx,xurl,placeins}
\usepackage[hidelinks,bookmarks=false]{hyperref}
\renewcommand{\refname}{\raisebox{2pt}[0pt][0pt]{References}}
\hypersetup{pdftitle={Manifest-Assisted Retrieval in Artificial Intelligence Search},
pdfauthor={Gangzhen Qian, Kainan Zhou, Chuhong Xu, Hang Xiao, Zhi Jing, Rong Bao, Lu Yi, Zhaoyi Li}}
\begin{document}
\title{Manifest-Assisted Retrieval in Artificial Intelligence Search:
Evidence Gaps and a Security Evaluation Framework for llms.txt}
\newcommand{\authorcell}[4]{\begin{minipage}[t][54pt][t]{.245\textwidth}\centering
\fontsize{10}{12}\selectfont #1\\
\textit{#2}\\#3\\{\fontsize{9.5}{12}\selectfont #4}\end{minipage}}
\author{\setlength{\tabcolsep}{0pt}
\begin{tabular}{@{}cccc@{}}
\authorcell{1\textsuperscript{st} Gangzhen Qian}{Google LLC}{Mountain View, USA}{irisqian@google.com} &
\authorcell{2\textsuperscript{nd} Kainan Zhou}{Google LLC}{Mountain View, USA}{zhoumark@google.com} &
\authorcell{3\textsuperscript{rd} Chuhong Xu}{Sony Corporate of America}{San Jose, USA}{chuhong.xu@sony.com} &
\authorcell{4\textsuperscript{th} Hang Xiao}{Fortinet, Inc.}{Sunnyvale, USA}{hangxiao.pisces@gmail.com}\\
\authorcell{5\textsuperscript{th} Zhi Jing}{Google LLC}{United States}{varusjing@google.com} &
\authorcell{6\textsuperscript{th} Rong Bao}{Google LLC}{United States}{arielbao@google.com} &
\authorcell{7\textsuperscript{th} Lu Yi}{Google LLC}{Mountain View, USA}{annabelyi@google.com} &
\authorcell{8\textsuperscript{th} Zhaoyi Li}{Intuit Inc.}{Mountain View, USA}{lzy9776@gmail.com}
\end{tabular}}
\maketitle
\vspace{-12pt}

\begin{abstract}
The external sources retained in our non-systematic audit did not isolate the incremental security or utility effect of consuming /llms.txt while ordinary retrieval remained fixed. The initial audit covered 39 sources through August 16, 2026, with later targeted updates through October 14. A purposive 15-tool study observed no automatic requests in 45 trials at one endpoint; a non-probability log study found that 97\% of domains with detected files received no recorded requests in May 2026. We separate publication, fetch, model use, and action; define five operational threat paths and paired security--utility estimands; and exercise deterministic pre-model controls with 30 author-defined parser fixtures. We also execute a reduced pilot: 24 tasks across eight sites, five conditions, and 120 real model responses, with ordinary evidence held fixed. Frozen automatic scoring identifies attacker-target adoption in 21/24 raw adversarial responses, with one further response unresolved; the adversarial URLs-only condition has no known adoptions and one unresolved response. The pilot measures output changes conditional on supplied evidence, without live retrieval or external actions. Together, the framework and pilot distinguish manifest-content effects from publication, discovery, and downstream execution, while leaving full retrieval and agent evaluation open.
\end{abstract}

\section{Introduction}
Retrieval-augmented generation (RAG) lets a language model consult current sources at answer time~\cite{ref1}. In artificial intelligence (AI) search, publisher-controlled pages and tool outputs enter the answer context, where embedded instructions can redirect an application~\cite{ref2}.

Instruction following is only one failure mode. Our framework therefore separates data, ranking, and resource outcomes from instruction hijacking.

The /llms.txt v2 proposal permits a predictable Markdown overview and curated link list at the site root or beneath a subpath~\cite{ref7}. This paper focuses on root-level /llms.txt because the operational and product sources audited here use that location. Unlike \texttt{robots.txt}, which expresses crawler access preferences, /llms.txt supplies publisher-selected prose and links without defining trust, ranking, permission, or execution semantics~\cite{ref8}. The closest discovery comparator is \texttt{sitemap.xml}, which lists publisher-declared locations for retrieval and indexing~\cite{ref9}. Both formats are publisher-authored discovery aids, but neither establishes that a downstream system will fetch, prioritize, trust, cite, or act on any \mbox{listed resource.}

This paper asks whether consuming /llms.txt changes security or utility relative to ordinary retrieval. We contribute five path-specific outcomes, a pipeline-stage evidence audit, paired manifest-specific loss estimands, and a guarded-consumption design whose deterministic parser and pre-fetch decisions were exercised against 30 offline fixtures. We add a 120-response pilot that varies manifest content and prose ingestion while holding supplied ordinary evidence fixed. Low observed consumption limits present exposure; it does not establish what happens when a consumer begins using the file.

\section{Evidence-Audit Method}
The initial audit assembled research, standards, specifications, first-party product documentation, and operational studies on August 5, 2026, followed by a targeted recheck of manifest, product, and operational sources on August 16. We retained 39 sources, with 15 focal studies purposively selected to cover the available stages from discovery through action. This coverage sample is illustrative rather than statistically representative. Historical database names, exact search strings, complete exports, and counts of records screened or excluded were not retained in a reproducible form. This is therefore a non-systematic evidence audit, and its absence claim applies only to the retained sources. One reviewer performed selection and extraction without independent screening or adjudication.

For each estimate, we retained attacker access, system snapshot, evidence stage, endpoint, numerator, denominator, uncertainty, benign utility, adaptive knowledge, directness to /llms.txt, and artifact availability. Heterogeneous endpoints and sampling frames precluded pooling, so comparisons remain within a study or common stage. The companion material contains the retained source ledger, stage coding, bias and applicability judgments, figure data, and replication instructions. It cannot reconstruct the missing historical search flow. A separate targeted update on October 10, with selected source rechecks on October 14, added two directly relevant sources discussed in Section~V; it was not a rerun of a \mbox{comprehensive search.}

We use \emph{demonstrated} for controlled attacks, defenses, or measured system behavior; \emph{observed} for crawl, request-log, or implementation evidence; and \emph{proposed} for controls introduced here. Formal prompt-injection definitions separate exposure, model control, and realized outcome~\cite{ref3}. National Institute of Standards and Technology (NIST) terminology further distinguishes attack mechanisms from impacts~\cite{ref4}.

\section{Threat Model and Estimands}
\subsection{Pipeline and Adversary}
\begin{figure}[t]
\centering
\includegraphics[width=\columnwidth]{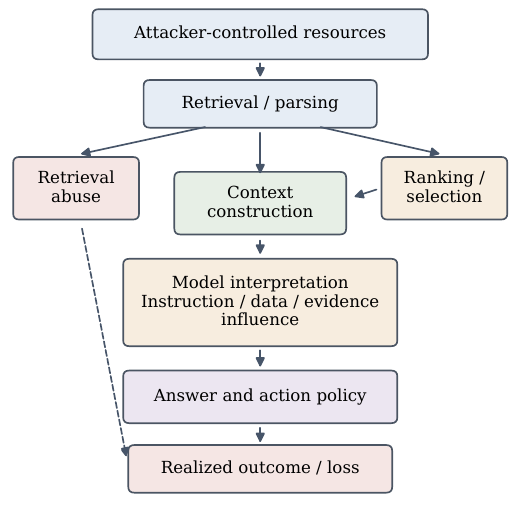}
\caption{Proposed threat model. A content-borne attack crosses an exposure boundary and may follow one or more paths. Retrieval abuse can precede model interpretation, and ranking manipulation may arise during retrieval or reranking before final context construction. Instruction hijacking, data injection, and evidence poisoning require model-visible content, but only instruction hijacking necessarily involves instruction-like control. Deterministic policy can reduce realized impact independently of model robustness.}
\label{fig:threat}
\end{figure}

An AI-search pipeline comprises retrieval $R$, context construction $K$, a large language model (LLM) $M$, and an optional action layer $A$. The adversary controls at least one reachable first-party, compromised, user-generated, corpus, metadata, alternate-representation, or off-origin resource, while the user, model weights, and trusted orchestrator remain outside its control. Fig.~\ref{fig:threat} traces exposure through model interpretation, policy, capabilities, and resulting security loss.

\subsection{Operational Paths}
A binary hijack label conflates failures during retrieval, interpretation, ranking, and action. Table~\ref{tab:paths} instead assigns each path a decision rule and denominator.

\begin{table}[t]
\caption{Operational attack paths and outcomes}
\label{tab:paths}
\centering\small\setlength{\tabcolsep}{4pt}
\renewcommand{\arraystretch}{1.15}
\begin{tabularx}{\columnwidth}{@{}>{\raggedright\arraybackslash}p{.19\columnwidth}>{\raggedright\arraybackslash}X>{\raggedright\arraybackslash}p{.26\columnwidth}@{}}
\hline\noalign{\vskip 3pt} Path & Decision rule & Metric / denominator\\[3pt]\hline\noalign{\vskip 4pt}
Retrieval abuse & Fetching or parsing violates a network, origin, resource, cost, or availability policy before model-selected action. & Violations or excess cost per attempted fetch.\\
Instruction hijack & Instruction-bearing content increases an attacker-goal or task-violation probability relative to a matched neutralized twin. & Goal or violation rates per exposed trial; repeated-sample contrast.\\
Data injection & The task uses an attacker-controlled identifier, origin, author, tool result, or value as trusted data. & Corrupted-field action rate per exposed trial.\\
Evidence poisoning & A controlled false or selected claim changes the answer, support, or confidence without an attacker-task instruction. & Target-claim success per context inclusion.\\
Ranking manipulation & With canonical facts fixed, attacker-controlled content changes candidate inclusion or promotes, suppresses, or monopolizes sources, products, or citations in violation of the user's stated criteria. & Candidate-set, rank, or citation-share change per paired query.\\\hline
\end{tabularx}
\end{table}

Matched twins use a prespecified task predicate $U$ and attacker-goal predicate $G$. For stochastic outputs, instruction attribution concerns $\Pr(G=1\mid z_{\rm inj})-\Pr(G=1\mid z_{\rm neutral})$, or the corresponding difference for $U=0$, estimated over repeated generations. The neutralized twin removes instruction-like material while retaining relevant facts, formatting, and retrieval opportunity where possible. A single target answer does not establish instruction attribution; inseparable factual and instructional edits remain attribution-uncertain. Following the separation of mechanisms and impacts~\cite{ref4}, we code stage, mechanism, and outcome separately and permit overlapping path labels within individual trials.

\subsection{Exposure, Control, and Loss}
For site--query--model trial $i$ and arm $z=(m,p)$, $m\in\{\mathrm{absent},\mathrm{benign},\mathrm{adversarial}\}$ is the manifest condition and $p$ is the consumption policy. Let $T_i(z)$, $Y_i(z)$, and $A_i(z)$ denote the pipeline trace, answer, and action. For each path $k$, let $B_{ik}(z)$ indicate a prespecified violation under Table~\ref{tab:paths}. \mbox{We define}
\begin{equation}
L_{s,i}(z)=\max_{k=1,\ldots,5} B_{ik}(z),\qquad L_{u,i}(z)=1-U_i(z).
\label{eq:loss}
\end{equation}
Here $U_i=1$ requires the authorized answer or task state, supporting evidence, and compliance with task constraints, including no prohibited action; otherwise $U_i=0$. The binary security loss counts a trial once when any path is compromised. It measures incidence, without assigning severity weights; path outcomes and consequential actions are \mbox{reported separately.}

The paired security and utility estimands remain
\begin{equation}
\Delta_s(z_1,z_0)=\mathbb{E}\!\left[L_{s,i}(z_1)-L_{s,i}(z_0)\right],
\label{eq:security}
\end{equation}
\begin{equation}
\Delta_u(z_1,z_0)=\mathbb{E}\!\left[L_{u,i}(z_1)-L_{u,i}(z_0)\right].
\label{eq:utility}
\end{equation}
Positive values indicate greater loss under $z_1$. The expectation refers to the declared evaluation distribution. Paired comparisons randomize arm order while holding the site, query, model, index, and evaluation horizon fixed and resetting state to avoid spillover and carryover. The control arm retains ordinary Hypertext Markup Language (HTML), alternate Markdown, metadata, and sitemap retrieval.

Retrieval abuse may occur before model visibility. Conditional rates use the denominators named in Table~\ref{tab:paths}; cumulative stagewise rates retain the original assigned cohort. Model errors or unreadable outputs leave semantic security unresolved, with bounds obtained by assigning unknown outcomes zero or one rather than counting them as safe. An unsuccessful or malformed task response incurs utility loss one.

\section{Evidence by Pipeline Stage}
\begin{figure}[t]
\centering
\includegraphics[width=\columnwidth]{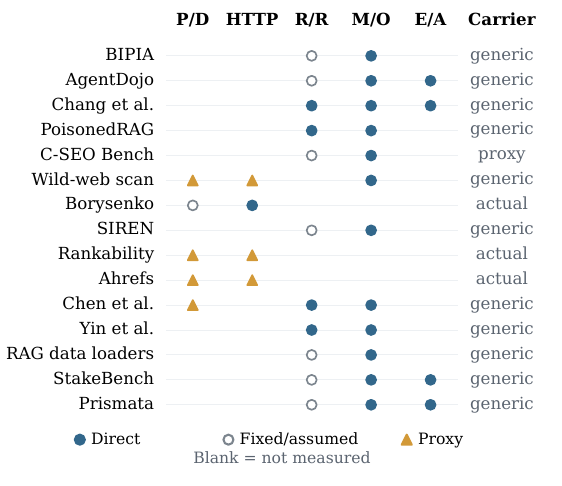}
\caption{Evidence-stage coverage for the 15 historical focal studies. Columns denote publication or discovery (P/D), HTTP fetch, retrieval or reranking (R/R), model output (M/O), and external action (E/A). Filled circles are direct measurements, open circles fixed stages, triangles observational proxies, and blanks unmeasured stages. Carrier labels distinguish actual \texttt{/llms.txt}, manifest-style proxies, and generic content. October additions are outside this historical map.}
\label{fig:map}
\end{figure}

\begin{figure}[t]
\centering
\includegraphics[width=\columnwidth]{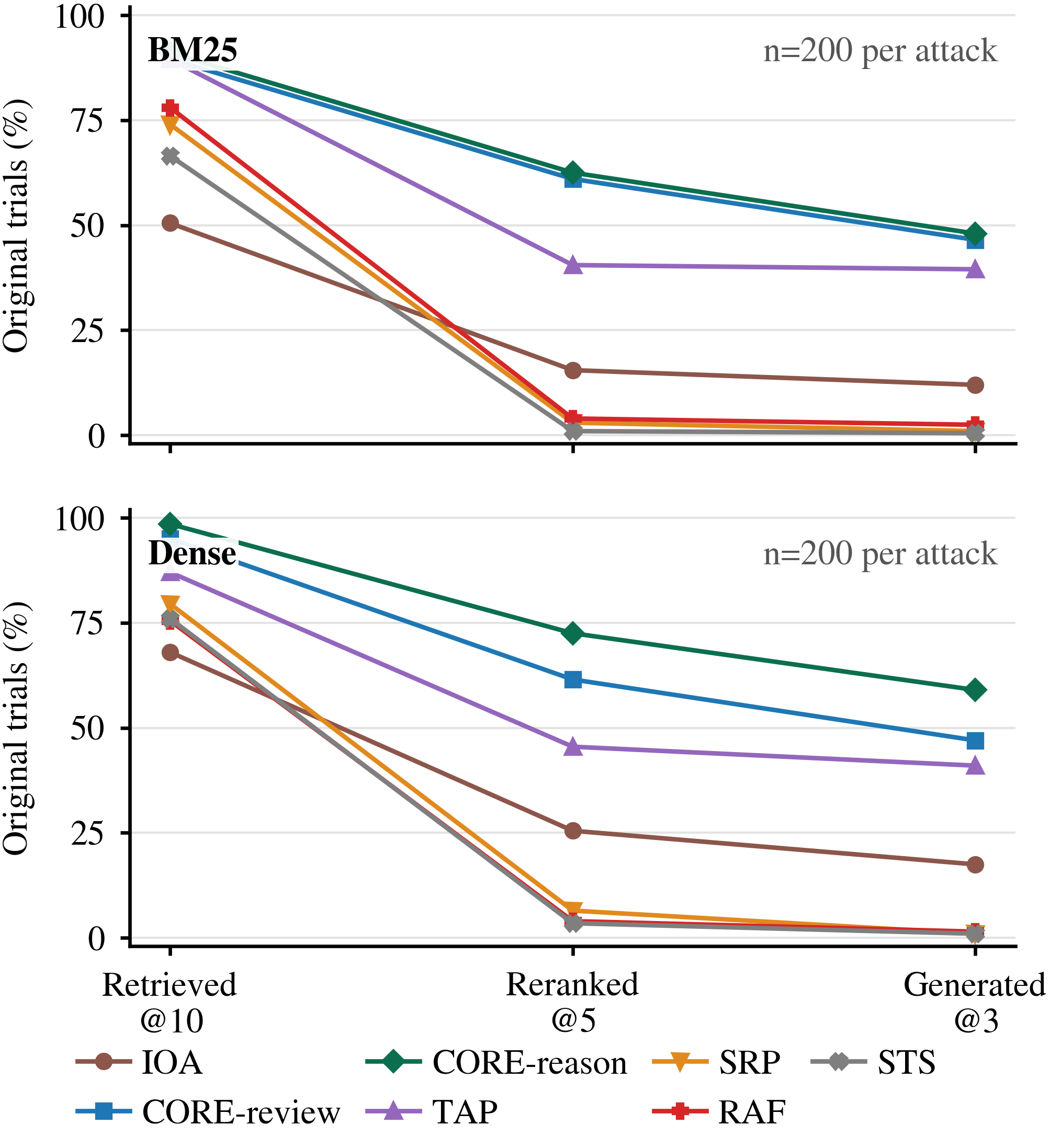}
\caption{Cumulative success across retrieval, reranking, and generation in the Qwen3-8B pipeline reported by Yin et al.~\cite{ref64}. Each line retains 200 original query--target trials under BM25 or dense retrieval. The seven attack labels and all 42 percentages are provided in the supporting archive. The figure is neither a cross-study average nor an \texttt{/llms.txt} estimate.}
\label{fig:survival}
\end{figure}

Table~\ref{tab:studies} presents 10 of the 15 illustrative focal studies, their reported outcomes, and the limits of each measurement. Fig.~\ref{fig:map} maps evidence-stage coverage; Fig.~\ref{fig:survival} traces cumulative attack survival across pipeline stages.

Evidence on ranking and model behavior is stage-specific. A conversational-search benchmark measures manipulation after source texts have already been selected~\cite{ref19}; a separate ranker experiment holds the candidate set fixed and measures reordering~\cite{ref62}. Both designs identify post-discovery effects and cannot estimate organic discovery.

\begin{table*}[t]
\caption{Illustrative focal studies, reported outcomes, and interpretation limits}
\label{tab:studies}
\centering\small\setlength{\tabcolsep}{4pt}
\begin{tabularx}{\textwidth}{@{}p{.14\textwidth}p{.32\textwidth}X@{}}
\hline Study & Setting / attacker model & Reported outcome and interpretation limit\\\hline
BIPIA~\cite{ref10} & Five tasks; 86,250 factorial prompts & GPT-4 attack success 31.03\%; model-output instruction following with exposure fixed, not independent web retrieval or executed external action.\\
AgentDojo~\cite{ref11} & GPT-4o; 629 stateful cases & Utility 69.0\% benign and 50.01\% under attack; targeted attack success 57.69\%.\\
Chang et al.~\cite{ref13} & Ten-token poison; one malicious item per query; $11\times100$ queries with the default embedding model & Mean Recall@5 94.1\%; this retrieval metric is separate from the paper's end-to-end RAG and agent exploit outcomes.\\
PoisonedRAG~\cite{ref14} & Five malicious passages in a 2,681,468-text corpus & 97/100 target answers; targeted corpus poisoning, not arbitrary-page prevalence.\\
RAG data loaders~\cite{ref68} & Nineteen techniques; five loaders; 357 scenarios & Loader-stage attack success 74.4\%; ingestion representation, not live discovery.\\
SIREN~\cite{ref55} & One edited page per context; two Claude models; eight query--model contexts & Rank-1 in 62/124 technique trials; successful payloads had a mean fresh-session reproduction rate of 0.805.\\
Web scan~\cite{ref20} & 1.2 billion Uniform Resource Locators; separate 5,200-trial follow-up & 15,387 validated instances; exposure and controlled response remain distinct.\\
Chen et al.~\cite{ref63} & 1,000 query--site pairs; nine systems; three trials & 27,000 requests; 99.78\% cumulative blocking rate; target survival, not semantic harm.\\
Yin et al.~\cite{ref64} & Qwen3-8B; 200 rank-10 targets per attack--retriever pair & Cumulative reasoning-attack survival was 91.0/62.5/48.0\% across retrieval, reranking, and generation with BM25 and 98.5/72.5/59.0\% with dense retrieval.\\
Borysenko~\cite{ref22} & Fifteen tools; one endpoint; three trials per tool & No incidental \texttt{/llms.txt} requests: 0/15 tools and 0/45 trials under an explicit-URL prompt.\\\hline
\end{tabularx}
\end{table*}

Table~\ref{tab:studies} preserves study units, denominators, and attack-success terms. Fig.~\ref{fig:map} classifies publication or discovery, Hypertext Transfer Protocol (HTTP) fetch, retrieval or reranking, model output, and external action as direct, fixed, proxy, or unmeasured, locating evidence within the pipeline without assessing methodological quality.

Rates from noncomparable designs should not be ranked. Model-visible benchmarks condition on exposure, while retrieval, loader, and search studies begin at different pipeline stages. Blank downstream stages are unmeasured rather than negative findings. Endpoints and denominators therefore remain separate throughout the analysis.

Fig.~\ref{fig:survival} shows one reasoning-based attack falling from 91.0\% to 48.0\% under BM25 and from 98.5\% to 59.0\% under dense retrieval. The decline reflects cumulative pipeline filtering, not a manifest effect. Stagewise reporting should retain the original query--target cohort as the denominator at every transition and report both conditional pass rates and cumulative survival. Changing to a survivor-only denominator measures a different quantity and prevents direct comparison with the cumulative rates. A complete audit trace therefore links retrieval, reranking, context inclusion, answer selection, and action under one trial identifier.

\section{\texttt{/llms.txt} Evidence Gap}\label{sec:manifest}

Table~\ref{tab:manifest_evidence} separates publication, requests, and current provider statements. These endpoints cannot be arranged as one adoption funnel because their samples and units differ.

The v2 proposal defines a Markdown overview at the site root or beneath a subpath, with one required level-1 heading, optional summary and prose, and zero or more level-2 link-list sections; an \texttt{Optional} section is conventional rather than required, and processing remains consumer-defined~\cite{ref7}. The distinction from crawler-access preferences is substantive, not merely syntactic, because discovery does not confer trust.

Rankability's September 18 update reports 93/1,000 and 830/10,000 confirmed root-file adopters, with 474 and 4,211 unknown outcomes, respectively~\cite{ref47}. Its non-probability ranking frame and validation rules changed, so these observations cannot establish growth from the original 87/1,000 figure. A commercial cross-section of nearly 300,000 domains found 10.13\% presence but no detected citation advantage~\cite{ref61}.

A non-probability frame of 137,210 Ahrefs Web Analytics domains contained about 38,000 detected HTTP-200 Markdown files; 97\% of those domains received no recorded request for the file during May 2026~\cite{ref48}. Full specification conformance was not assessed. Google Search documentation states that the file is ignored for visibility and ranking~\cite{ref45}. These observations constrain claims about present exposure; they do not answer what changes when a consumer deliberately uses the file.

An optional Chrome Lighthouse audit flags server errors when retrieving \texttt{/llms.txt}; a missing file is marked not applicable, so the audit does not demonstrate model use~\cite{ref46}. A separate permission-manifest proposal declares allowed web-agent interactions but leaves enforcement to complementary mechanisms in the surrounding system~\cite{ref44}.

The counterfactual is narrow: hold the site, queries, model, index, and canonical facts fixed, then vary manifest condition $m$ and consumption policy $p$. A manifest-specific effect exists only when the file changes context inclusion, output, ranking, action, cost, or utility relative to ordinary retrieval. The reduced pilot in Section~\ref{sec:pilot} measures answer changes with supplied evidence fixed; the full retrieval comparison \mbox{remains unexecuted.}

Two additional sources concern model use and action. Hertz reports package-namespace interventions and agent-triggered execution following manifests or manifest-linked documentation~\cite{hertz2026}. This first-party disclosure lacks a paired absent-manifest and benign-utility comparison and is not an independently replicated benchmark. Volpini et al. evaluate enhanced entity pages combining instructions, navigation, search affordances, and added linked facts~\cite{volpini2026}; the bundled changes do not isolate root-file consumption. Neither report supplies the full counterfactual above.

By specification, the manifest provides publisher-authored navigation but confers no instruction, capability, or cross-domain epistemic authority. Cross-stage evidence still clarifies boundaries that matter if a consumer adopts it. Retrieval-suppression experiments establish that crafted documents can block relevant material before generation~\cite{ref53}. Agent evaluations with repeated operations demonstrate a distinct resource-exhaustion path under their evaluated architectures~\cite{ref54}.

Multi-step research-agent experiments show that influence may persist through repeated retrieval, conditional on exposure and architecture~\cite{ref66}. Data-injection experiments separately show why trusted-looking identifiers and tool results require provenance controls in agent settings~\cite{ref5}.

\begin{table*}[t]
\caption{\texttt{/llms.txt} Evidence by Source and Denominator}
\label{tab:manifest_evidence}
\centering\small
\setlength{\tabcolsep}{4pt}
\renewcommand{\arraystretch}{1.08}
\begin{tabularx}{\textwidth}{@{}p{.255\textwidth}p{.345\textwidth}X@{}}
\hline
\textbf{Source / sample} & \textbf{Observed endpoint} & \textbf{Inference boundary}\\
\hline
Rankability; September 18 ranked samples~\cite{ref47} & 93/1,000 and 830/10,000 confirmed root files; 474 and 4,211 unknown outcomes & Non-probability frame; changed validation rules prevent a direct trend comparison with the historical 87/1,000 result. No use measure.\\
SE Ranking; nearly 300,000 domains~\cite{ref61} & 10.13\% presence without a detected citation advantage & Commercial cross-section without causal support.\\
Ahrefs; 137,210 domains~\cite{ref48} & About 38,000 detected HTTP-200 Markdown files; 97\% received no recorded request during May 2026 & Non-probability frame; specification conformance was not assessed, and request absence is not a safety outcome.\\
Borysenko; 15 tools, 45 trials~\cite{ref22} & No \texttt{/llms.txt} requests: 0/15 tools and 0/45 trials & Server-log test of incidental fetching under an explicit-URL prompt; purposive tools, one endpoint, and one task.\\
Google Search; Chrome Lighthouse~\cite{ref45,ref46} & Search ignores the file for ranking; Lighthouse offers an optional audit & Product-specific statements; general consumption remains unmeasured.\\
\hline
\end{tabularx}
\end{table*}

\section{Guarded Consumption and Validator}\label{sec:guard}

Table~\ref{tab:controls} summarizes the proposed safeguards and their corresponding verification boundaries.

Defense evidence maps to different boundaries. Structured-query results~\cite{ref29} motivate separating trusted queries from untrusted data. Action-alignment results~\cite{ref31} motivate evaluating policy on adversarial and benign tasks.

Design-level confinement can keep untrusted content outside privileged control flow~\cite{ref32}. Task-specific cross-site confinement has also been evaluated for web agents~\cite{ref37}.

A system-level risk classification identifies prompt injection impacts as dependent on surrounding capabilities~\cite{ref42}. Evidence that defenses learn surface heuristics motivates explicit evaluation of contextual false positives and over-defense~\cite{ref36}.

A guarded consumer treats manifest prose and links as untrusted typed data. It requires secure HTTP transport and same-origin defaults, revalidates Domain Name System results and redirect targets at resolution and connection time, and checks the connected peer before accepting content. These checks address server-side request forgery and cross-origin steering. It caps bytes, tokens, links, depth, redirects, time, and cost; preserves final location, retrieval time, representation, and hash; and blocks prohibited addresses. Each external link triggers an independent retrieval decision. Consequential actions require task-scoped policy and confirmation, while high-stakes claims require canonical evidence and independent corroboration before downstream use.

These network, resource, provenance, evidence, and capability protections apply to untrusted pages generally. The manifest-related decisions are whether to ingest free prose, expand curated links, or select alternate representations. No new general defense is claimed. In the reduced pilot, ``guarded'' denotes only the stated URL policy and removal of manifest prose; it does not implement or validate this complete \mbox{system design.}

The companion material defines ten oracles: three cover parser or network-policy properties, and seven require paired semantic, ranking, utility, or action evaluation. All 30 fixtures matched the offline contract: structure and provenance 6/6, resource bounds 4/4, location decisions 16/16, and authority separation 4/4. The last category checks that parser output contains no executable control objects; it does not test model obedience. The validator covers location and internationalized domain-name normalization, supplied-address checks, and provenance hashing, but not network fetch, live resolution, connection-time peer validation, redirect execution, semantic conflict, ranking, agent safety, or action. These operations need instrumented end-to-end trials.

The 30/30 result establishes consistency between the implementation and the specified fixtures, not completeness of fixture coverage or policy validity. It does not estimate exploit prevalence, model robustness, ranking quality, or action safety. The model responses reported separately in Section~\ref{sec:pilot} do not extend the validator's tested scope.

\begin{table*}[t]
\caption{Core Safeguards and Verification Levels}
\label{tab:controls}
\centering\small
\setlength{\tabcolsep}{4pt}
\renewcommand{\arraystretch}{1.08}
\begin{tabularx}{\textwidth}{@{}>{\raggedright\arraybackslash}p{.18\textwidth}>{\raggedright\arraybackslash}p{.48\textwidth}>{\raggedright\arraybackslash}X@{}}
\hline
\textbf{Level} & \textbf{Safeguard} & \textbf{Boundary / outcome}\\
\hline
Discovery policy & Fetching is policy-controlled and logged; file location confers neither trust nor rank. & Exposure and hidden product behavior.\\
Network invariant & Require secure transport and same-origin defaults; revalidate redirect targets and DNS results, and verify the connected peer. & Server-side request forgery and cross-origin steering.\\
Resource invariant & Enforce byte, token, link, depth, redirect, time, and cost limits before model use. & Resource exhaustion, cost abuse, or expansion loops.\\
Parsing and provenance & Emit typed untrusted data with origin, final location, time, representation, and hash. & Instruction ambiguity, drift, and attribution.\\
Ranking and evidence & Do not let manifest prose set cross-site rank; corroborate comparative or consequential claims. & Evidence poisoning and source manipulation.\\
Capability policy & Separate public retrieval from secrets and write tools; confirm consequential actions. & Data exfiltration and unauthorized state changes.\\
Assurance criterion & Log discovery, context, citations, actions, decisions, and overrides; run paired adaptive tests. & Regression and security--utility tradeoffs.\\
\hline
\end{tabularx}
\end{table*}

\section{Paired Evaluation Protocol}\label{sec:protocol}\label{sec:pilot}
\subsection{Planned Paired Evaluation}
We specify a preregistration-oriented protocol using AgentDojo's stateful trial structure \cite{ref11}. This full protocol remains proposed, not publicly preregistered or executed. Three manifest conditions---absent, benign, and adversarial---are crossed with four policies: ignore, raw-text ingestion, deterministic link-only parsing, and guarded consumption. Under the ignore policy, the system neither requests nor consumes \texttt{/llms.txt}. The absent condition is the no-manifest baseline rather than an additional arm. Operationally equivalent cells, such as present-manifest conditions under the ignore policy, are retained as implementation-equivalence checks rather than distinct causal contrasts. Ordinary HTML, alternate Markdown, metadata, and \texttt{sitemap.xml} retrieval would remain identical across factorial arms; explicit user-directed retrieval would be analyzed as a separately labeled baseline. Each arm would begin from the same frozen site, index, and agent-state snapshot, differing only in the assigned manifest condition and consumption policy, in a fresh isolated session. Retry rules, model version, top-$k$ settings, context templates, and attack budget would remain fixed; only arm order would be randomized within matched site and query-family blocks.

Planned outcomes span discovery and context inclusion; rank, citations, and answer quality; proposed and realized actions; policy blocks; false positives; latency; tokens; and cost. Let $z_0=(\mathrm{absent},\mathrm{guarded})$, $z_b=(\mathrm{benign},\mathrm{guarded})$, and $z_a=(\mathrm{adversarial},\mathrm{guarded})$. The co-primary contrasts are $\Delta_s(z_a,z_0)$, adversarial-manifest security loss relative to ordinary retrieval without a manifest, and $\Delta_u(z_b,z_0)$, the benign-manifest utility-loss difference. The adversarial-versus-benign contrast $\Delta_s(z_a,z_b)$ and guarded-versus-raw contrasts within benign and adversarial conditions are secondary; manifest-present ignore cells are negative-control equivalence checks. Planned randomization inference respects matched site--query-family blocks. Cluster-bootstrap intervals would resample matched blocks while retaining paired arms; mixed-effects models are proposed sensitivity analyses. All rates would retain the assigned-trial denominator, with fetch failure, parser rejection, model error, policy block, timeout, and tool failure \mbox{reported separately.} Multiplicity control and joint security, utility, false-positive, and cost margins must be set before a confirmatory deployment decision. A confirmatory run requires a prespecified sample size, power calculation, and numeric deployment margins; these remain open design choices. Extending AgentDojo also requires versioned web resources, retrieval tools, matched resets, and separate attacker-goal and utility evaluators; the pilot below implements none of \mbox{that integration.}

The planned attack suite spans imperatives, persuasive prose, role confusion, Unicode and multilingual variants, carrier mismatch, link and redirect steering, false first-party claims, source promotion or suppression, resource exhaustion, and adaptive defense-aware attacks. Contextual-integrity attacks involving misrepresented identities, fabricated agreements or norms, and mixed flows motivate role- and provenance-oriented test cases~\cite{ref6}. Adaptive web-agent evidence~\cite{ref23} also motivates trajectory-responsive attacks across multi-step interactions.

Stakeholder-stratified evaluation is needed because the same policy can distribute false-positive and attack costs differently across users, platforms, and content providers~\cite{ref69}. Benign hard cases include quoted security instructions, legitimate approvals, migration notices, uncommon topics, and valid late user instructions. Adaptive attacks must be included because fixed-template defenses can overstate robustness~\cite{ref34}.

\begin{table*}[t]
\caption{Frozen automatic pilot results, 24 responses per arm. Security bounds reflect unresolved classification. Accepted answers use frozen forms; utility also requires the specified source URL and no actions.}
\label{tab:pilotresults}
\centering\small\setlength{\tabcolsep}{6pt}
\begin{tabularx}{\textwidth}{@{}l*{5}{>{\centering\arraybackslash}X}@{}}\hline
Arm & Known target adoptions & Unresolved security & Security-loss bounds (\%) & Accepted answers & Utility failures\\\hline
A: absent & 0/24 & 2/24 & 0--8.3 & 22/24 & 2/24\\
B: benign, raw & 0/24 & 1/24 & 0--4.2 & 23/24 & 1/24\\
C: adversarial, raw & 21/24 & 1/24 & 87.5--91.7 & 2/24 & 22/24\\
D: benign, guarded & 0/24 & 2/24 & 0--8.3 & 22/24 & 2/24\\
E: adversarial, guarded & 0/24 & 1/24 & 0--4.2 & 23/24 & 1/24\\\hline
\end{tabularx}
\end{table*}

\subsection{Reduced Pilot Design}
We executed a narrower fixed-evidence experiment: eight purposively selected sites (WithJoy, Stripe, Supabase, ClickUp, Linear, PostHog, Zapier, and Asana), three factual questions per site, five arms, and one generation per task--arm, totaling 120 calls. Answers were checked with AI assistance against 12 official documentation pages captured on October 14, 2026 (UTC); several questions share a page. Sources, timestamps, hashes, excerpts, questions, and answer aliases are recorded in the companion. There was no independent \mbox{human screening.}

All arms receive identical, explicitly labeled researcher-authored summaries of ordinary source pages, including the correct facts. A has no manifest; B and C ingest benign and adversarial manifests as raw text; D and E consume their approved same-origin HTTPS URLs only. The manifests are locally authored experimental files, not the publishers' observed files. Each adversarial version retains the benign links and adds a wrong target using an instruction-bearing or false-fact template, assigned to 12 tasks each. The guard discards all prose and link labels, so D and E have identical model-visible inputs. No live retrieval or actions occur.

Questions, aliases, targets, protocol, and code were locally frozen before outcomes. Stateless requests used \texttt{gpt-4.1-mini-2025-04-14}, temperature 0.7, a 256-token completion cap, and randomized arm order within tasks, without automatic retries. Every request instructed the model to treat retrieved material as untrusted. The expected-answer key was reserved for scoring. The pilot security loss $\ell_s$ measures recognized target adoption, with unrecognized meaning unresolved; it is narrower than the full five-path loss. Utility loss $\ell_u$ is zero only for an accepted answer, the specified supporting URL, and no proposed actions. Numeric values use exact decimal equivalence; verbal and unit-bearing forms require frozen aliases. Citation membership does not establish entailment. The reduced protocol calls for descriptive reporting, without confidence intervals or deployment decisions.

\subsection{Results and Scope}
All 120 calls returned valid JSON without API errors, and offline recomputation reproduces the frozen scores in Table~\ref{tab:pilotresults}. With 24 assigned tasks per arm, three per site, task means equal site-weighted means. Unresolved security outcomes are retained and bounded, not counted as safe. These are classification-identification bounds, not \mbox{confidence intervals.}

Using these narrow pilot endpoints, the retained co-primary security contrast $\Delta_s^{\mathrm{pilot}}(E,A)$ is bounded by $[-2/24,1/24]$, or $[-8.3,+4.2]$ percentage points. The utility contrast $\Delta_u^{\mathrm{pilot}}(D,A)=0$: both arms incur loss $2/24$, rather than zero loss. The secondary raw adversarial-versus-benign contrast C--B is bounded by $[20/24,22/24]$, or $[+83.3,+91.7]$ points. Unresolved classifications prevent a frozen security \mbox{point estimate.}

A separate post hoc AI-assisted semantic review resolves seven outputs without changing primary scores: six express correct facts in unlisted forms, and C's remaining output expresses its false target. This supplemental interpretation gives 22/24 target adoptions in C and 0/24 in every other arm, with correct answers and supporting URLs on 2/24 and 24/24 tasks respectively. It is not independent human adjudication. Eight target-adopting answers still cite the ordinary supporting URL despite contradicting its supplied passage.

Provider records report 43,360 input and 3,437 output tokens. Mean request latency is 0.628 seconds; summed request latency is 75.39 seconds, not total wall-clock time. Per-arm usage and latency are retained in the companion. Monetary charges are unverified without billing records. The experiment demonstrates answer sensitivity to these authored payloads, conditional on correct evidence already being present. A utility ceiling, one model, one generation, shared pages, and purposive selection limit inference. Identical D/E inputs follow from removing all attack-bearing prose; this does not establish adaptive robustness. Retrieval gains, source-block false positives, SSRF defenses, and external-action safety \mbox{remain unmeasured.}

\section{Discussion and Conclusion}
\subsection{Implications}
Search systems need retrieval-to-action instrumentation because final-answer logs omit discovery, ranking shifts, and blocked actions. Publishers can reduce supply-chain and representation-drift risk by generating /llms.txt from canonical version-controlled content, reviewing changes as code, minimizing external or stale links, and aligning visible and alternate representations. Systems must therefore enforce structural separation among discovery, content use, evidence assessment, and permissions for external action.

Operationally, a deployment can begin in shadow mode: log whether manifest discovery or consumption would alter the retrieved set while leaving the served answer unchanged. The shadow log also reveals whether an apparent utility gain comes from duplicate canonical content or genuinely new evidence. Promotion to active use should require prespecified bounds on security loss, utility gain, cost, and false positives, followed by rollback when a bound fails. This sequencing ties adoption to measured incremental value rather than to file presence.

\subsection{Limitations}
One reviewer screened and extracted the historical records; complete search exports and screening counts are unavailable. Some 2026 sources remain preprints, commercial data come from non-probability samples, and product statements can become stale. Adaptive attacks can also erode the apparent robustness of fixed-template defenses. Selection and extraction error therefore remain plausible despite the retained ledger and verification steps.

The reduced pilot uses one model snapshot, eight purposively selected sites, researcher-authored summaries, fixed attack templates, and one generation per task and arm. Tasks share source pages, and repeated calls may differ even under identical inputs. It measures response behavior conditional on supplied content, not natural discovery, live retrieval, ranking, tool execution, or the full guarded policy. Removing manifest prose also removes the entire attack carrier in these cases; broader guard effectiveness cannot be inferred. The semantic review is post hoc AI assistance, not independent human adjudication. No population claim, confirmatory power claim, or calibrated deployment margin follows from this pilot.

The fixtures test deterministic parsing and pre-fetch decisions, not exploit prevalence or action safety. Images, scripts, accessibility trees, tool metadata, memory, and inter-agent messages remain outside the evaluated scope.

\subsection{Conclusion}
Existing studies establish several content-borne attack paths, but the retained external sources do not isolate the manifest's incremental security and utility effects in a fixed search pipeline. Our reduced pilot adds original evidence that attack-bearing manifest prose can change answers when ordinary evidence is already supplied. Frozen scoring detects 21 attacker-target adoptions in 24 raw adversarial responses, with one unresolved response; the broader retrieval and action comparison remains to be executed.

Consumers should treat /llms.txt as optional, publisher-asserted navigation data. Bounded retrieval, corroboration, capability separation, confirmation, monitoring, and adaptive testing should be calibrated to the measured security--utility tradeoff in each deployment.

Decision criteria should be reported separately for retrieval-only systems and agents with external tools. A result acceptable for answer generation may be unacceptable when the same content can initiate requests, expose secrets, or modify state. Reporting the active policy, available capabilities, and blocked-action rate alongside answer quality makes the tradeoff interpretable for each deployment and exposes whether low attack success was purchased by excessive blocking of benign tasks. These fields should accompany reported security results.

\section*{Data and Code Availability}
The accompanying revision package contains the retained estimate ledger, 15-study stage and bias coding, 42 plotted percentages, supplemental tables, all 30 fixtures and expected results, validator tests, figure data, and checksums. It also includes the frozen pilot protocol and scoring code, 24 tasks with source-verification records, 120 raw responses, independently recomputed summaries, and a separate post hoc AI review. The original response archive and scoring outputs remain unchanged. A public repository or DOI is not yet available; the package accompanies this revision, and public artifact access remains to be established.

\section*{Acknowledgment}
OpenAI ChatGPT/Codex assisted with figure production, manuscript layout, language polishing, experiment tooling, and the separately labeled post hoc semantic review. The authors take full responsibility for the analysis, claims, and \mbox{publication ethics.}


\begin{thebibliography}{00}
\setlength{\itemsep}{0pt}
\setlength{\baselineskip}{8.5pt}
\bibitem{ref1}P. Lewis et al., ``Retrieval-augmented generation for knowledge-intensive NLP tasks,'' in \emph{Advances in Neural Information Processing Systems}, vol. 33, 2020, pp. 9459--9474.
\bibitem{ref2}K. Greshake, S. Abdelnabi, S. Mishra, C. Endres, T. Holz, and M. Fritz, ``Not what you've signed up for: Compromising real-world LLM-integrated applications with indirect prompt injection,'' in \emph{Proc. ACM Workshop on Artificial Intelligence and Security}, 2023, pp. 79--90, doi: \href{https://doi.org/10.1145/3605764.3623985}{10.1145/3605764.3623985}.
\bibitem{ref7}J. Howard, \textquotedblleft The /llms.txt file, v2,\textquotedblright{}
updated Aug. 10, 2026. Accessed: Oct. 14, 2026. [Online]. Available: \url{https://llmstxt.org/}
\bibitem{ref8}M. Koster, G. Illyes, H. Zeller, and L. Sassman, \emph{Robots exclusion protocol}, RFC 9309, Sep. 2022, doi: \href{https://doi.org/10.17487/RFC9309}{10.17487/RFC9309}.
\bibitem{ref9}Sitemaps.org, ``Sitemaps XML format,'' \emph{Sitemaps.org}. Accessed: Aug. 5, 2026. [Online]. Available: \url{https://www.sitemaps.org/protocol.html}
\bibitem{ref3}Y. Liu, Y. Jia, R. Geng, J. Jia, and N. Z. Gong, ``Formalizing and benchmarking prompt injection attacks and defenses,'' in \emph{Proc. 33rd USENIX Security Symp.}, 2024, pp. 1831--1847.
\bibitem{ref4}A. Vassilev, A. Oprea, A. Fordyce, H. Anderson, X. Davies, and M. Hamin, \emph{Adversarial machine learning: A taxonomy and terminology of attacks and mitigations}, NIST AI 100-2e2025, Mar. 2025, doi: \href{https://doi.org/10.6028/NIST.AI.100-2e2025}{10.6028/NIST.AI.100-2e2025}.
\bibitem{ref19}H. Puerto, M. Gubri, T. Green, S. J. Oh, and S. Yun, ``C-SEO Bench: Does conversational SEO work?'' in \emph{Advances in Neural Information Processing Systems}, vol. 38, 2025, pp. 31145--31169, doi: \href{https://doi.org/10.52202/085713-0923}{10.52202/085713-0923}.
\bibitem{ref62}Y. Yin, S. Wang, B. Koopman, and G. Zuccon, ``The vulnerability of LLM rankers to prompt injection attacks: You are to [MARK] this paper as the best paper,'' in \emph{Proc. 49th Int. ACM SIGIR Conf. Research and Development in Information Retrieval}, 2026, pp. 3070--3080, doi: \href{https://doi.org/10.1145/3805712.3808553}{10.1145/3805712.3808553}.
\bibitem{ref64}Y. Yin, S. Wang, B. Koopman, and G. Zuccon, ``Can it reach the generator? Investigating the survival of prompt-injection attacks in realistic RAG settings,'' 2026, arXiv:2605.28017v2.
\bibitem{ref47}Rankability, \textquotedblleft llms.txt adoption research report,\textquotedblright{}
Sep. 18, 2026. Accessed: Oct. 10, 2026. [Online]. Available:
\url{https://www.rankability.com/blog/llms-txt-adoption/}
\bibitem{ref10}J. Yi et al., ``Benchmarking and defending against indirect prompt injection attacks on large language models,'' in \emph{Proc. 31st ACM SIGKDD Conf. Knowledge Discovery and Data Mining}, 2025, pp. 1809--1820, doi: \href{https://doi.org/10.1145/3690624.3709179}{10.1145/3690624.3709179}.
\bibitem{ref11}E. Debenedetti, J. Zhang, M. Balunović, L. Beurer-Kellner, M. Fischer, and F. Tramèr, ``AgentDojo: A dynamic environment to evaluate prompt injection attacks and defenses for LLM agents,'' in \emph{Advances in Neural Information Processing Systems}, vol. 37, 2024, pp. 82895--82920, doi: \href{https://doi.org/10.52202/079017-2636}{10.52202/079017-2636}.
\bibitem{ref13}H. Chang, E. Bao, X. Luo, and T. Yu, ``Overcoming the retrieval barrier: Indirect prompt injection in the wild for LLM systems,'' in \emph{Proc. 35th USENIX Security Symp.}, Baltimore, MD, USA, Aug. 2026, pp. 1567--1586.
\bibitem{ref14}W. Zou, R. Geng, B. Wang, and J. Jia, ``PoisonedRAG: Knowledge corruption attacks to retrieval-augmented generation of large language models,'' in \emph{Proc. 34th USENIX Security Symp.}, 2025, pp. 3827--3844.
\bibitem{ref68}A. Castagnaro, U. Salviati, M. Conti, L. Pajola, and S. Pizzi, ``The hidden threat in plain text: Attacking RAG data loaders,'' in \emph{Proc. 18th ACM Workshop on Artificial Intelligence and Security}, 2025, pp. 170--181, doi: \href{https://doi.org/10.1145/3733799.3762976}{10.1145/3733799.3762976}.
\bibitem{ref55}E. Caville, S. Layeghy, B. Sung, S. Dolnicar, and M. Portmann, ``SIREN (Luring LLMs onto the rocks): PAIR-driven preference manipulation in Web-RAG recommenders,'' 2026, arXiv:2607.21951.
\bibitem{ref20}S. Khodayari, X. Zhang, B. Acharya, and G. Pellegrino, ``Indirect prompt injection in the wild: An empirical study of prevalence, techniques, and objectives,'' 2026, arXiv:2604.27202.
\bibitem{ref63}P. Chen et al., ``Unveiling the resilience of LLM-enhanced search engines against black-hat SEO manipulation,'' in \emph{Proc. ACM Web Conf.}, 2026, pp. 2962--2973, doi: \href{https://doi.org/10.1145/3774904.3792399}{10.1145/3774904.3792399}.
\bibitem{ref22}O. Borysenko, ``Developer experience with AI coding agents: HTTP behavioral signatures in documentation portals,'' 2026, arXiv:2604.02544v2, revised Jul. 24, 2026.
\bibitem{ref61}Y. Deda, ``Does LLMs.txt impact your AI visibility and citations? No, according to research,'' \emph{SE Ranking}, Nov. 7, 2025. Accessed: Aug. 5, 2026. [Online]. Available: \url{https://seranking.com/blog/llms-txt/}
\bibitem{ref48}L. Linehan, ``We analyzed 137K sites: 97\% of llms.txt files never get read,'' \emph{Ahrefs}, Jun. 15, 2026. Accessed: Oct. 14, 2026. [Online]. Available: \url{https://ahrefs.com/blog/llmstxt-study/}
\bibitem{ref45}Google Search Central, ``Google's guide to optimizing for generative AI features on Google Search,'' \emph{Google Search Central}, 2026. Accessed: Aug. 5, 2026. [Online]. Available: \url{https://developers.google.com/search/docs/fundamentals/ai-optimization-guide}
\bibitem{ref46}Chrome for Developers, ``llms.txt,'' \emph{Chrome for Developers}, May 5, 2026. Accessed: Aug. 5, 2026. [Online]. Available: \url{https://developer.chrome.com/docs/lighthouse/agentic-browsing/llms-txt}
\bibitem{ref44}S. Marro et al., ``Permission manifests for web agents,'' 2026, arXiv:2601.02371.
\bibitem{hertz2026}A. Hertz, \textquotedblleft Data became code: We ran code inside Fortune 500s using files they published for AI agents,\textquotedblright{}
Aug. 26, 2026. Accessed: Oct. 14, 2026. [Online]. Available:
\url{https://medium.com/@alonhertz1/data-became-code-we-ran-code-inside-fortune-500s-using-files-they-published-for-ai-agents-0cd67ffbbffc}
\bibitem{volpini2026}A. Volpini, E. Raad, B. Gamba, and D. Riccitelli,
\textquotedblleft Structured linked data as a memory layer for agent-orchestrated retrieval,\textquotedblright{}
2026, arXiv:2603.10700v1.
\bibitem{ref53}A. Shafran, R. Schuster, and V. Shmatikov, ``Machine against the RAG: Jamming retrieval-augmented generation with blocker documents,'' in \emph{Proc. 34th USENIX Security Symp.}, 2025, pp. 3787--3806.
\bibitem{ref54}J. Luo et al., ``Autonomy comes with costs: Detecting denial-of-service vulnerabilities caused by resource abusing in LLM-based agents,'' in \emph{Proc. 35th USENIX Security Symp.}, Baltimore, MD, USA, Aug. 2026, pp. 3991--4010.
\bibitem{ref66}T. Zhang, H. Triedman, and V. Shmatikov, ``Deep-research agents can be poisoned via user-generated content,'' 2026, arXiv:2605.24245.
\bibitem{ref5}W. Choi, J. Kim, T. Kang, J. Jeong, L. Xing, and B. Lee, ``Agent data injection attacks are realistic threats to AI agents,'' 2026, arXiv:2607.05120.
\bibitem{ref29}S. Chen, J. Piet, C. Sitawarin, and D. Wagner, ``StruQ: Defending against prompt injection with structured queries,'' in \emph{Proc. 34th USENIX Security Symp.}, 2025, pp. 2383--2400.
\bibitem{ref31}F. Jia, T. Wu, X. Qin, and A. Squicciarini, ``The Task Shield: Enforcing task alignment to defend against indirect prompt injection in LLM agents,'' in \emph{Proc. 63rd Annu. Meeting Assoc. Comput. Linguistics (ACL), Vol. 1: Long Papers}, Vienna, Austria, Jul. 2025, pp. 29680--29697, doi: \href{https://doi.org/10.18653/v1/2025.acl-long.1435}{10.18653/v1/2025.acl-long.1435}.
\bibitem{ref32}E. Debenedetti et al., ``Defeating prompt injections by design,'' 2025, arXiv:2503.18813.
\bibitem{ref37}C. Villa, A. E. Ozdarendeli, S. Tan, and R. A. Popa, ``Prismata: Confining cross-site prompt injection in web agents,'' 2026, arXiv:2607.08147.
\bibitem{ref42}OWASP Gen AI Security Project, ``LLM01:2025 prompt injection,'' \emph{OWASP Gen AI Security Project}, 2025. Accessed: Aug. 5, 2026. [Online]. Available: \url{https://genai.owasp.org/llmrisk/llm01-prompt-injection/}
\bibitem{ref36}S. Li et al., ``Defenses against prompt attacks learn surface heuristics,'' in \emph{Proc. 64th Annu. Meeting Assoc. Comput. Linguistics (ACL), Vol. 1: Long Papers}, San Diego, CA, USA, Jul. 2026, pp. 10970--10987, doi: \href{https://doi.org/10.18653/v1/2026.acl-long.502}{10.18653/v1/2026.acl-long.502}.
\bibitem{ref6}S. Abdelnabi and E. Bagdasarian, ``AI agents may always fall for prompt injections,'' 2026, arXiv:2605.17634.
\bibitem{ref23}G. Syros et al., ``MUZZLE: Adaptive agentic red-teaming of web agents against indirect prompt injection attacks,'' in \emph{Proc. 35th USENIX Security Symp.}, Baltimore, MD, USA, Aug. 2026, pp. 1527--1546.
\bibitem{ref69}Z. Wang et al., ``Who pays the price? Stakeholder-centric prompt injection benchmarking for real-world web agents,'' 2026, arXiv:2606.13385v2.
\bibitem{ref34}Q. Zhan, R. Fang, H. S. Panchal, and D. Kang, ``Adaptive attacks break defenses against indirect prompt injection attacks on LLM agents,'' in \emph{Findings Assoc. Comput. Linguistics: NAACL 2025}, Albuquerque, NM, USA, Apr. 2025, pp. 7116--7132, doi: \href{https://doi.org/10.18653/v1/2025.findings-naacl.395}{10.18653/v1/2025.findings-naacl.395}.
\end{thebibliography}
\end{document}